\documentclass[pre,twocolumn,showpacs,amsmath,amssymb]{revtex4}

\usepackage{graphicx}
\usepackage{wasysym}
\usepackage{dcolumn}
\usepackage{bm}

\begin{document}

\title{Linear and anomalous front propagation in system
with non Gaussian diffusion: the importance of tails}

\author{Maurizio Serva$^{1}$, Davide Vergni$^{2}$, 
Angelo Vulpiani$^3$}

\affiliation{$^1$Dipartimento di Ingegneria e Scienze dell'Informazione e Matematica, Universit\`a dell'Aquila, L'Aquila, Italy}

\affiliation{$^2$Istituto per le Applicazioni del Calcolo ``Mauro Picone'' - CNR, Roma, Italy}

\affiliation{$^3$Dipartimento di Fisica, Universit\`a di Roma "Sapienza", and ISC-CNR, Roma, Italy}

\date{\today}

\begin{abstract}
We investigate front propagation in systems with diffusive and
sub-diffusive behavior.  The scaling behavior of moments of the
diffusive problem, both in the standard and in the anomalous cases, is
not enough to determine the features of the reactive front. In fact,
the shape of the bulk of the probability distribution of the transport
process, which determines the diffusive properties, is important just
for pre-asymptotic behavior of front propagation, while the precise
shape of the tails of the probability distribution determines
asymptotic behavior of front propagation.
\end{abstract}


\maketitle

\section{Introduction}

Reaction-diffusion processes appear in a large class of phenomena from
combustion to ecology~\cite{neufeld2010}.  In presence of non trivial
geometry (e.g. graphs)~\cite{burioni2005} or anomalous diffusion
processes~\cite{Barkai2014,Klafter2015} the reaction dynamics is an
intriguing and difficult issue~\cite{mancinelli2002, mancinelli2003,
burioni2012}.

In particular, the case of reaction in subdiffusive 
systems~\cite{Sokolov2006,Mendez2009,Fedotov2010,Volpert2013} is
rather interesting for its role in living systems, since
anomalous subdiffusion has been reported for different biological
transport problems such as two-dimensional diffusion in the plasma
membrane and three-dimensional diffusion in the nucleus and cytoplasm
(see~\cite{Saxton2007} and reference therein).

The aim of the present work is to show that the behavior of tails of
the probability distribution of the pure transport problem has
the most important role for the front propagation properties, which can
be anomalous even in case of standard diffusion. On the other hand
one can have the standard linear front propagation also in presence of
subdiffusion. It is important to stress that, as already noted 
in~\cite{Sokolov2006,Mendez2009} and clearly stated 
in~\cite{Nepomnyashchy2016}, the details of the reaction-diffusion
rules can be very important for the front propagation features.

The paper is organized as follows: in Sect.~\ref{sec:2} we briefly
summarize some results about non standard diffusive systems and front
propagation in presence of reaction.  Sect.~\ref{sec:3} is devoted to
an analysis of front propagation in a simplified model with
sub-diffusion.  In Sect.~\ref{sec:4} we show that, even in a more
realistic model, the basic ingredient for the asymptotic behavior of
front propagation is given by the shape of the tail of the probability
distribution, $P(x,t)$, of the pure transport problem. We point out
that the shape of the bulk of $P(x,t)$, which determines the diffusive
properties, could be important just for the pre-asymptotic behavior
of the front propagation. Sect.~\ref{sec:5} is dedicated to the
conclusions.

\section{A survey of known results\label{sec:2}}
Let us indicate with $P(x,t)$ the probability distribution
(or the probability density) that a walker is 
in $x$ at time $t$. 
Under rather general hypothesis, $P(x,t)$ satisfies the equation
\begin{equation}
P(x,t+\Delta t) = \int_{-\infty}^{+\infty} P(x-u,t)\rho_{\Delta t}(u,x,t)du,
\label{P}
\end{equation}
where $\rho_{\Delta t}(u,x,t)$ is the probability density to be in $x$
at time $t+\Delta t$ under the condition to be in $x-u$ at time
$t$. Even if $\rho_{\Delta t}(u,x,t)$ has an explicit dependence on
$x$ and $t$ the process is Markovian.  When $\rho_{\Delta t}(u,x,t)$
does not depend on $x$ and $t$ and in absence of fat tails, Eq.~(\ref{P}) 
describes standard diffusion, i.e. at large time $\langle
x^2(t)\rangle\sim t$ and $P(x,t)$ is close to a Gaussian. 
The simplest case is the standard random walk, where $\Delta t=1$ and
$\rho(u)=\frac{1}{2}(\delta(u-1)+\delta(u+1))$.

In order to introduce a reaction term in the diffusion
equation~(\ref{P}) we consider~\cite{mancinelli2002} a time discrete
reaction process such as, in absence of diffusion, $\theta(x, t+\Delta
t)=G_{\Delta t}(\theta(x, t))$, where $\theta(x,t)$ is the
concentration field, and $G_{\Delta t}$, the reaction map, can be
approximated by $G_{\Delta t}(\theta)=\theta + \frac{\Delta t}{\tau}
f(\theta)$ where $f(\theta)$ is the reaction term which, in the simple
auto-catalytic case, reads $f(\theta)=\theta(1-\theta)$ and $\tau$ is
the characteristic time associated to the reaction.  In any case, we
are interested in pulled reactions (i.e., $f''(\theta)<0$ and
$f'(0)=1$, as in the case of auto-catalytic reactions) for which the
detailed shape of $f(\theta)$ is not important.

Following~\cite{mancinelli2002} the concentration evolution
satisfies the equation
\begin{equation}
   \theta(x,t+\Delta t) = \int_{-\infty}^{+\infty} 
	\rho_{\Delta t}(u,x,t) G_{\Delta t}(\theta(x-u,t)) {\mathrm d}u \,.
   \label{integ}
\end{equation}
The above equation extends, for general diffusive
process~(\ref{P}) in a discrete time version, the usual
reaction-diffusion equation $\partial_t\theta=D\nabla
\theta+f(\theta)/\tau$.\\ For equation~(\ref{integ}) it is possible to
apply the maximum principle~\cite{Freidlin,mancinelli2003} obtaining
\begin{equation}
\theta(x,t) \leq P(x,t) \,e^{{\displaystyle{\max_\theta}} \{f(\theta)/\theta\} t / \tau}\,.
\label{approxgen}
\end{equation}

In the case of pulled reaction, to which we always refer in this
paper, the previous equation becomes (see~\cite{mancinelli2005})
\begin{equation}
\theta(x,t) \sim P(x,t) \,e^{t / \tau}\,.
\label{approx}
\end{equation}
For a standard diffusive process, i.e. when 
$P(x,t) \sim e^{-x^2 / 4Dt}$, Eq.~(\ref{approx}) reads
\begin{equation}
\theta(x,t) \sim e^{-x^2 / 4Dt+ t / \tau},
\label{thetaP}
\end{equation}
where $D$ is the diffusion coefficient.  The front position
$x_F(t)$ can be easily obtained assuming that the concentration
$\theta(x,t)$, as it is given by Eq.~(\ref{thetaP}), 
is of order 1.  It turns out:
\begin{equation}
x_F(t)=2\sqrt{\frac{D}{\tau}} \, t,
\label{xf}
\end{equation}
representing the well-known result of a linear propagating front with
velocity $v_f=2\sqrt{D/\tau}$.

In the case of anomalous diffusion, a simple assumption for the
probability distribution that generalizes the Gaussian shape
for the standard diffusive case, is 
\begin{equation}
P(x,t) \sim e^{-C(x /  t^{\nu})^{\beta}}.
\label{Pnu}
\end{equation}
The value of $\nu$ discriminates between subdiffusion and
superdiffusion, with $\nu< 1/2$ and $\nu>1/2$, respectively.
Eq.~(\ref{Pnu}), in fact, implies that
\begin{equation}
\langle x^2(t) \rangle \sim t^{2 \nu}.
\label{x2}
\end{equation}
Using Eq.~(\ref{Pnu}) in Eq.~(\ref{approx}) one gets
\begin{equation}
\theta(x,t) \sim e^{-C(x / t^{\nu})^{\beta} + t / \tau }\,,
\label{thetaP2}
\end{equation}
so that, as previously discussed, one can easily obtain the front
position $x_F(t)$ that, in this case, behaves as
\begin{equation}
x_F(t) \sim t^{\delta} \, , \,\,\,\,\,\, \delta=\nu+
\frac{1}{\beta}.
\label{xf2}
\end{equation}
Let us stress that, also when (\ref{Pnu}) holds, the
value of the exponent $\delta$ depends on the characteristics of the process
underlying anomalous diffusion. An interesting case, suggested
by an argument due to Fisher~\cite{Fisher, mancinelli2003}, is when
\begin{equation}
\beta=\frac{1 }{ 1- \nu}\,.
\label{beta}
\end{equation}
Using expression (\ref{beta}) in Eq.~(\ref{xf2}) one gets that the
front always propagates linearly in time, i.e. $\delta=1$ although
diffusion is anomalous ($\nu \neq 1/2$).  Such a linear behavior holds
both in the sub-diffusive case (for example in the random walk on a comb
lattice where $\nu=1/4$) and in the super-diffusive one (for example
in the random shear flow where
$\nu=3/4$)~\cite{mancinelli2003}.

In general, when relations (\ref{Pnu}) or (\ref{beta}) do not hold,
the front propagation could be non linear in time.  We refer
to these cases as ``non standard'' propagation.  We will see that non
standard propagation can occurs also in the case of a standard
diffusive process when the tail of the $P(x,t)$ show a non standard
behavior.\\
As an example we report the work of~\cite{mancinelli2002} where the
diffusion process is given by (\ref{P}) but the probability density of
jumps $\rho(u)$ has fat tails, i.e, $\rho(u) \sim 1/|u|^{\alpha+1}$
with $\alpha$ positive.  The central limit theorem ensures that for
large times $t$ and for $\alpha \ge 2$
\begin{equation}
\frac{|x(t)|}{t^{1/2}}=
\frac{1}{t^{1/2}} \sum_{s=1}^t u_s  \to \omega
\label{lim}
\end{equation}
where the $u_s$ are independent extractions with probability $\rho(u)$ and
$\omega$ is a Gaussian variable.  
Accordingly, the standard diffusion scaling
\begin{equation}
\langle |x(t)|^q\rangle \sim t^{q/2},
\label{av}
\end{equation}
holds (provided that $q<\alpha$ in order to avoid
divergence induced by tails).  Therefore one would expect that the
associate propagating front scales linearly with time, but, on the
contrary, one finds that its behavior is exponential in time. In fact,
while the bulk of the $P(x,t)$ becomes Gaussian when $t$ increases,
the tails continue to be fat as $1/|x|^{\alpha+1}$, although the
frontiers between the Gaussian bulk and the fat tails shift at the
extremes when time increases.  Using Eq.~(\ref{approx}) one obtains
\begin{equation}
\theta(x,t) \sim \frac{1}{|x|^{\alpha+1}} \, e^{t / \tau},
\label{theta3}
\end{equation}
so that for large times the front increases exponentially 
fast~\cite{mancinelli2002}:
\begin{equation}
 x_F(t) \sim  e^{t / (\alpha+1)\tau}.
\label{xf3}
\end{equation}
In conclusion, the shape of the tails and not the scaling of 
$\langle |x(t)|^q\rangle$ is a crucial ingredient in determining
the features of the front propagation. 

\section{Diffusion and sub-diffusion in a simple model\label{sec:3}}
In the previous Section we have seen that, in presence of fat
tails of the probability distribution for the pure transport process,
standard diffusion are compatible with super-linear behavior for the
front propagation.  The main goal of the present paper is to confirm
that the shape of such tails is the most important element influencing
the scaling behavior of the front position. In particular we are
interested in sub-diffusive and (more importantly) standard diffusive
system when slim tails of the probability distribution induce to a
sub-linear behavior of the front propagation.  In this respect, the
present paper is complementary to the result presented at the end of
the previous section and in Mancinelli et al.~\cite{mancinelli2002}.

In order to tackle this goal we consider a very simple random walk
model which, according to a single control parameter, can be
sub-diffusive or diffusive: a walker, starting from home ($x(0)=0$),
at any discrete time can make a (unitary length) step to the right or
to the left
\begin{equation}
x(t+1)=x(t)+\sigma(t),
\label{xx}
\end{equation}
where
\begin{itemize}
\item
$\sigma(t)= \pm 1$ with equal probability if $|x(t)| \le t^\lambda$;
\item
$\sigma(t)= -{\rm sign}(x(t))$ if $|x(t)| > t^\lambda$.
\end{itemize}

\begin{figure}
\includegraphics[scale=0.65]{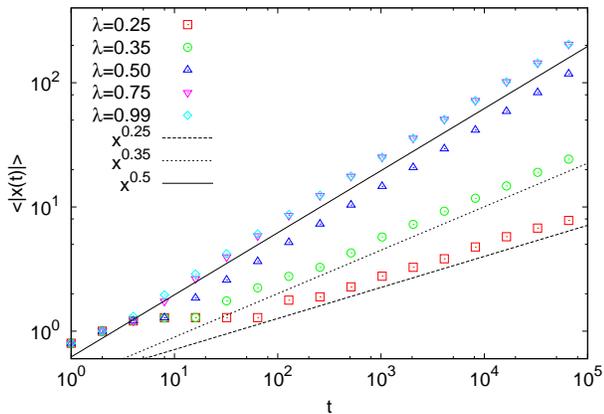} 
\caption{Diffusion. The random walk process (\ref{xx}) is sub-diffusive 
when $0 < \lambda < 1/2$, according to $\langle|x(t)| \rangle \sim
t^\nu$ with $\nu=\lambda$, and is standard diffusive $\langle|x(t)|
\rangle \sim t^{1/2}$ when $1/2 \le \lambda < 1$.}
\label{Fig_1}
\end{figure}

\begin{figure}
\includegraphics[scale=0.65]{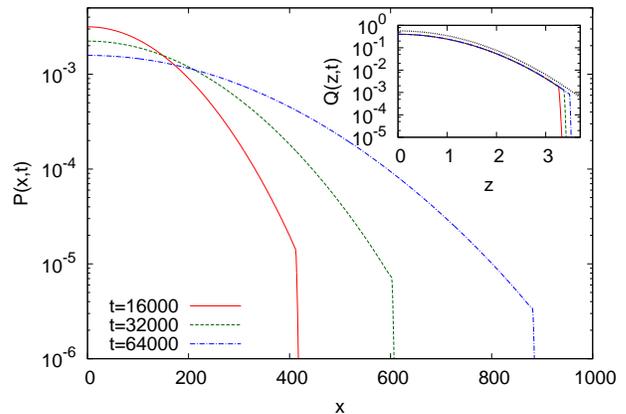}
\caption{Probability. The right side of the probability distribution 
$P(x,t)$ of the process (\ref{xx}) is shown for $\lambda=0.55$ and
different times. Although the value of $\gamma$ is close to $1/2$ the
truncation only involves distribution tails which carry a
minimal total probability.  Moreover, truncation
shifts to the right in time and asymptotically vanishes.  This can be
better appreciated in the inset where the probability $Q(z,t)$ of
$z(t)=x(t)/t^{1/2}$ is shown and compared with a Gaussian 
(dotted black line).}
\label{Fig_2}
\end{figure}

This process is confined in a box by two reflecting boundaries
situated in $L$ and $-L$ where $L=t^\lambda$.

If $0 < \lambda < 1/2$ the process is sub-diffusive according to
$\langle|x(t)| \rangle \sim t^\nu$ with $\nu=\lambda$, as shown in
Fig.~\ref{Fig_1}.  Notice that in this case the distribution of
$z(t)=x(t)/t^\nu$ is approximately uniform between -1 and 1 (for large
times) as the box dimension increases slowly with respect to the
relaxation of the process in the box.

On the contrary, if $1/2 \le \lambda \le 1$ the process behaves as an
ordinary diffusion with $\langle|x(t)| \rangle \sim t^{1/2}$ so that
$\nu=1/2\neq\lambda$, as shown in
Fig.~\ref{Fig_1}.  In this ordinary diffusive region the effect of the
reflecting boundaries is just to kill the tails of the process for
$|x| > t^\lambda$.  The distribution of $z(t)=x(t)/t^{1/2}$, as time
$t$ increases, becomes closer to a Gaussian, with the truncation which
shifts on the extremes and disappears for large times $t \to \infty$.
This is shown in Fig.~\ref{Fig_2}.

\begin{figure}
\includegraphics[scale=0.65]{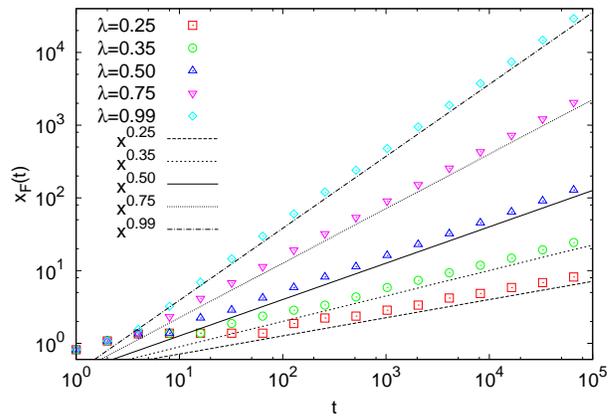}
\caption{Reaction. Front position $x_F$ for different values
of $\lambda$ and $\tau=1$ is shown. In all cases 
$x_F(t) \sim t^\delta$ with $\delta = \lambda$.}
\label{Fig_3}
\end{figure}

Model (\ref{xx}) is a simplified version of the non Markovian
random walk introduced in Ref.s~\cite{serva2013,serva2014}.  
The stochastic process (\ref{xx}) is a Markovian process 
so we can use Eq.(2) to describe reaction-diffusion dynamics.

When a reaction term is involved, 
it turns out that in all cases the exponent $\delta$ of the
propagating front equals the exponent $\lambda$ which defines the
process, i.e. $x_F(t) \sim t^\delta$ with $\delta = \lambda$.
Therefore, not only the scaling behavior of the front is anomalous in
case of sub-diffusion (for $0<\lambda<1/2)$, but also in case of
regular diffusion (for $1/2 \le \lambda < 1)$.  In fact, e.g.
in the last case, Eq.~(\ref{approx}) reads:
\begin{equation}
\theta(x,t) \sim e^{-x^2 / 4Dt}\Theta(t^\delta - |x|) e^{t / \tau}\,,
\label{theta4}
\end{equation}
where $\Theta(x)$ is the Heaviside step function representing
the truncation of the distribution tails. Since the truncation
overcomes the linear propagation due to the Gaussian diffusion
together with the exponential growing of the reaction,
one has $x_F(t)\sim t^\delta$, as can be
appreciated in Fig.~\ref{Fig_3} where the front position $x_F(t) =
\min\{ \, |x| \, ; \, \theta(x,t)<0.5 \}$ is shown. However,
completely equivalent results is obtained by computing the
average position of reaction products or their total quantity.  

\section{The role of tails in a model with subdiffusion \label{sec:4}}
We now consider a more complex model that includes the previously
neglected tails, in order to investigate their effect on propagation.
We consider
\begin{equation}
x(t+1)=x(t)+\sigma(t),
\label{xxx}
\end{equation}
with $\sigma(t)$ given by
\begin{itemize}
\item
$\sigma(t)= {\rm sign}(x(t))$ with probability $p(x(t),t)$ and
\item
$\sigma(t)= -{\rm sign}(x(t))$ with probability $1-p(x(t),t)$; 
\end{itemize}
where
\begin{equation}
p(x,t)=\frac{1}{2+\left(\frac{|x|}{t^\lambda}\right)^\eta}.
\label{xpx}
\end{equation}
\begin{figure}
\includegraphics[scale=0.65]{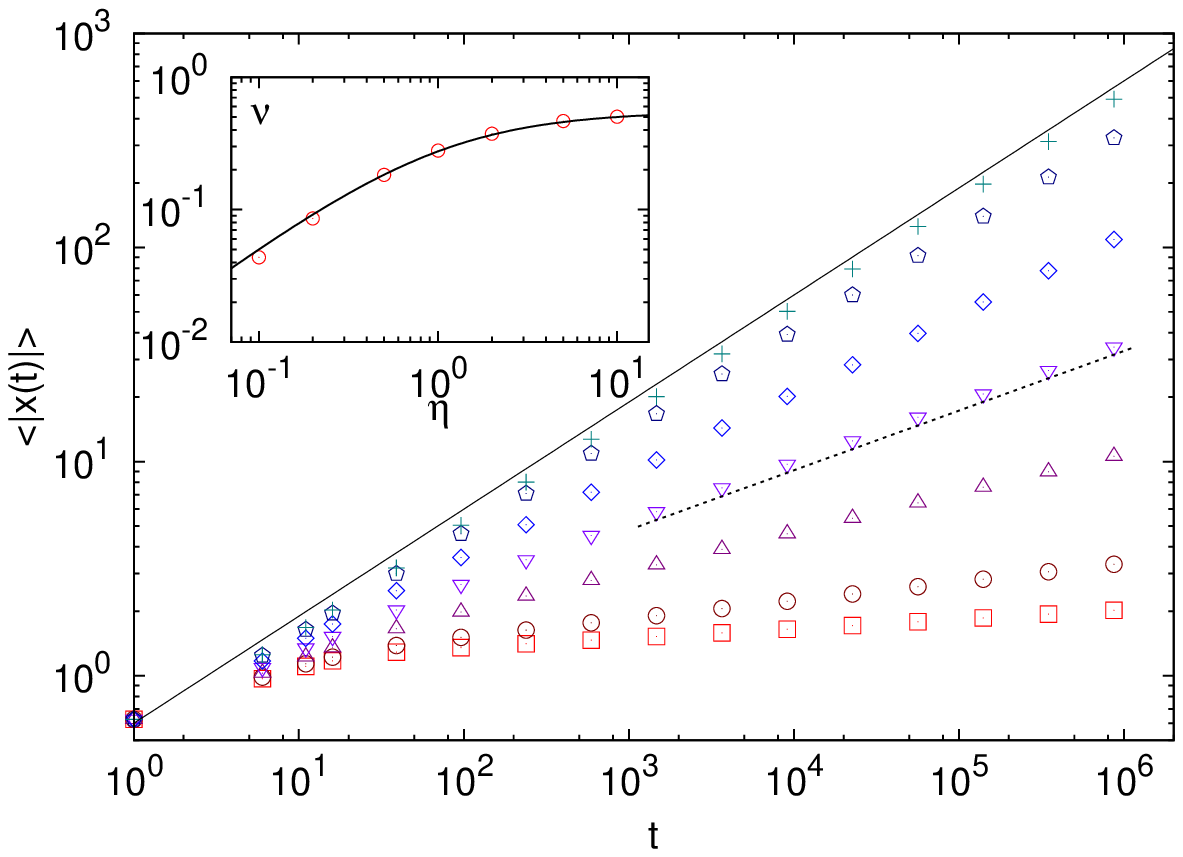}
 \caption{Diffusion. The average  $\langle|x(t)| \rangle$ is shown for 
$\lambda=0.55$ for different values of $\eta$
(from top to bottom 
$\eta = 10\,\,(+)$, 
$\eta = 5.0\,\,(\pentagon)$,
$\eta = 2.0\,\,(\diamond)$, 
$\eta = 1.0\,\,(\bigtriangledown)$,
$\eta = 0.5\,\,(\bigtriangleup)$, 
$\eta = 0.2\,\,(\circ)$, 
$\eta = 0.1\,\,(\Box)$).
The dashed line is the scaling behavior $\langle|x(t)| \rangle\sim
t^\nu$ for $\eta=1$, with $\nu$ given by Eq.~(\ref{scalingnu}).  In the
inset the measured scaling $\nu$ is reported ($\circ$) together with
the prediction (\ref{scalingnu}) (solid line) at varying $\eta$.}
\label{Fig_4}
\end{figure}
The above model has been motivated by ecological problems.  
For instance the problem of foraging strategies, with the walker
(animal) changing its attitude when it is at the frontier of
unexplored regions~\cite{serva2013,serva2014}.

Such a process does not present sharp boundaries.  In fact, if $\eta$
is finite, the walker can cross the position $|x|=t^\lambda$ but she
advances with an increasing difficulty when she is at a larger
distance from the origin.  However, in the limit $\eta \to \infty$,
this model coincides with the model of Section III (which has
boundaries in $|x|=t^\lambda$).  In the opposite case, i.e. for $\eta
\to 0$, it is worth noting that the process becomes an
``asymmetrical'' random walk with probabilities $1/3$ for $\sigma(t)=
{\rm sign}(x(t))$ and $2/3$ for $\sigma(t)= -{\rm sign}(x(t))$ (in the
origin the probability is symmetrical).  In this case the walker does
not diffuse anymore and the stationary probability $P(x)$ can be
easily computed from detailed balance finding $P(x)=\frac{1}{3}\,
2^{-|x|}$.
\\
The average $\langle|x(t)| \rangle$ is shown in Fig.~\ref{Fig_4} for
$\lambda=0.55$ and different values of $\eta$.  
As a consequence of the previous discussion for $\eta$ small the
diffusion decline (and for $\eta\to 0$ the diffusion is absent,
with $\langle |x(t)| \rangle = $ const.) while for $\eta$ large
the process tends to standard diffusion ($\langle |x(t)|
\rangle \sim t^{1/2}$ since $\lambda \ge 1/2$).  For intermediate 
values of $\eta$, the process is always subdiffusive.
A simple argument to compute the
anomalous exponent is as follows.  Using Eq.~(\ref{xxx}) 
and writing
$\langle x(t)\sigma(t)\rangle$ in terms of the conditional expected
value with respect to the probability~(\ref{xpx}) one obtains the
exact evolution rule for $\langle x^2(t) \rangle$
\begin{equation}
\langle x^2(t+1) \rangle - \langle x^2(t) \rangle = 1 -
\left\langle \frac{2 |x(t)| \left(\frac{|x(t)|}{t^\lambda}\right)^\eta}
{2+\left(\frac{|x(t)|}{t^\lambda}\right)^\eta} \right\rangle
\label{xquadro}
\end{equation}
First of all we assume a diffusive dynamics,
$\langle |x(t)| \rangle\sim t^\nu$, with monofractal properties, i.e.
$$\langle |x(t)|^\eta 
\rangle \sim \langle |x(t)| \rangle^\eta \sim t^{\nu \eta}\,.$$
Such a conjecture has been checked a posteriori.
We observe that forcely $\nu < \lambda$, because in the opposite case
the r.h.s. of Eq.~(\ref{xquadro}) becomes negative, leading to a
non-diffusive dynamics.  Therefore one can safely conclude that
$\left(|x(t)|/t^\lambda\right)^\eta$ is vanishing for large times,
and Eq.~(\ref{xquadro}) can be approximated by
\begin{equation}
\langle x^2(t+1) \rangle - \langle x^2(t) \rangle \sim 1 -
\frac{1}{t^{\lambda \eta}} \langle |x(t)|^{\eta+1} \rangle\,.
\label{approxxquadro}
\end{equation}
The above expression can be analyzed considering three different cases:
\begin{description}
\item[$\nu > \lambda \eta / (1 + \eta)$] : the second 
term of equation (\ref{approxxquadro}) is negative implying
absence of diffusion and contradicting the above assumptions.
\item[$\nu = \lambda \eta / (1 + \eta)$] :  
equation (\ref{approxxquadro}) is coherent as far as $\lambda
\eta /(1 + \eta) \leq 1/2$, since the maximum scaling exponent consistent
with equation (\ref{approxxquadro}) is $\nu=1/2$.
\item[$\nu < \lambda \eta / (1 + \eta)$] : in this case 
the second term of equation (\ref{approxxquadro}) is equal to one (for
large times) implying $\nu=1/2$.
\end{description}
In conclusion:
\begin{equation}
\nu = \min \left[ \frac{1}{2}, \; \frac{\lambda \eta}{1 + \eta} \right]\,
\label{scalingnu}
\end{equation}
and, in the case of $0 < \lambda \le 1/2$ one always has $\nu =\lambda
\eta /(1 + \eta)$.  Notice that, for any $0 < \lambda \le 1$ 
and in the limit $\eta \to \infty$, one recovers
the result already found for the model with sharp boundaries.
In the inset of Fig.~\ref{Fig_4} the perfect agreement of
the prediction~(\ref{scalingnu}) with the anomalous diffusive
exponent $\langle |x(t)| \rangle\sim t^\nu$ is shown.

For the understanding of the features of front propagation
we have to determine the shape of the probability
distribution, $P(x,t)$, of the process (\ref{xxx}). Accordingly to 
the definition of the process we have the following conditional
expectation with respect to $x(t)=x$:
\begin{equation}
\langle \sigma(t)|x \rangle = -{\rm sign}(x)
\frac{\left(\frac{|x|}{t^\lambda}\right)^\eta}
{2+\left(\frac{|x|}{t^\lambda}\right)^\eta}.
\end{equation}
Let us use the standard procedure to continuously approximate the
forward Kolmogorov equation (Fokker-Planck equation in the terminology
used in physics) for the probability density $P=P(x,t)$ as
\begin{equation}
\frac{\partial P}{\partial t}= -\frac{\partial (b P)}{\partial x} +
\frac{1}{2} \frac{\partial^2 (a P)}{\partial x^2} 
\end{equation}
where 
\begin{equation}
b(x,t)=\langle \sigma(t)|x \rangle = -{\rm sign}(x)
\frac{\left(\frac{|x|}{t^\lambda}\right)^\eta}
{2+\left(\frac{|x|}{t^\lambda}\right)^\eta}.
\label{defb}
\end{equation}
and
\begin{equation}
a(x,t)= \langle \sigma^2 (t)|x \rangle - 
        \langle \sigma(t)|x \rangle^2 = 1-b^2(x,t).
\label{defa}
\end{equation}
\begin{figure}
\includegraphics[scale=0.65]{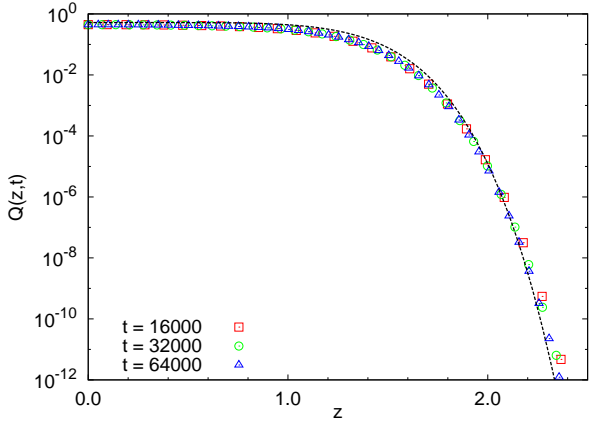}
\caption{Probability. The right side of the probability distribution 
$P(x,t)$ of the diffusive process~(\ref{xxx}) is shown using the rescaled
variable $z(t)=x(t)/t^\nu$ with $\nu$ given by Eq.~(\ref{scalingnu})
for $\lambda=0.55$ and $\eta=5.0$ at different times.  The dashed
line is the rescaled distribution $Q(z)=\exp(-z^{\eta+1}/(\eta+1))$ as
from Eq.~(\ref{predP}).}
\label{Fig_5}
\end{figure}
\begin{figure}
\includegraphics[scale=0.65]{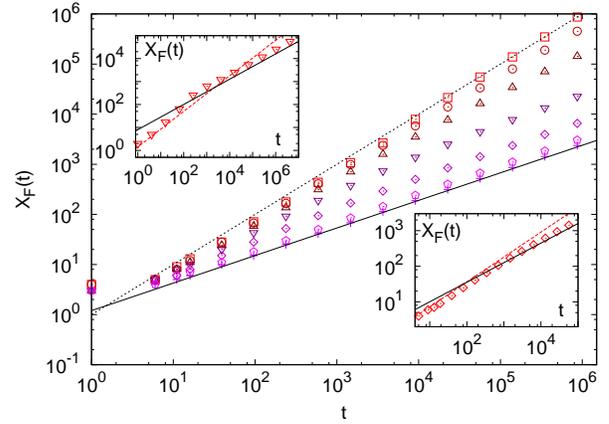}
\caption{Reaction. Front position for $\lambda=0.55$,
with reaction parameter $\tau=1$ and different values of $\eta$ (from
bottom to top, $\eta = 10\,\,(+)$, $\eta = 5.0\,\,(\pentagon)$, $\eta
= 2.0\,\,(\diamond)$, $\eta = 1.0\,\,(\bigtriangledown)$, $\eta =
0.5\,\,(\bigtriangleup)$, $\eta = 0.2\,\,(\circ)$, $\eta =
0.1\,\,(\Box)$) is shown.  In all cases the scaling is $x_F(t) \sim
t^\delta$ with $\delta=\lambda=0.55$ (solid line) but there is a
transient in which a different exponent (see Eq.~(\ref{scalingdelta}))
is selected. In the limit case $\eta\to 0$ a linear front behavior
($x_F(t) \sim t$ dashed line) is restored.  In the inset the scaling
behavior in the transient region $x_F(t)=(1+\eta)^{1/(1+\eta)}
t^\delta$ with $\delta$ given by Eq.~(\ref{scalingdelta}) (dashed
line) is shown together with the asymptotic behavior
$x_F(t)=(A)^{1/\eta} t^\lambda$, with $A=8$ corresponding to a
threshold $1/10$ for the jump probability~(\ref{xpx}) (solid line) for
$\eta=1.0$ (top left) and for $\eta=2.0$ (bottom right).  The matching
between the two scalings gives the crossover time (see
Eq.~(\ref{eq:match})).}
\label{Fig_6}
\end{figure}
Moreover, with the change of variable $y=x/t^\nu$ one can define the 
probability density for $y$ as  
$\tilde{P} (y,t) dy = P (y \, t^\nu, t)  dx $. Since $dx= t^\nu \, dy$, 
one has $\tilde{P} (y,t) = P (y \, t^\nu, t) t^\nu $.
This density satisfies the following equation: 
\begin{equation}
\frac{\partial \tilde{P}}{\partial t}=
\frac{\nu }{t} \frac{\partial (y  \tilde{ P})}{\partial y} -
\frac{1}{t^\nu} \frac{\partial (\tilde{b} \tilde{P})}{\partial y} +
\frac{1}{2t^{2\nu}}  \frac{\partial^2 (\tilde{a} \tilde{P})}{\partial y^2} 
\label{ap1eqd}
\end{equation}
where $\tilde{a}(y,t)= a(y \, t^\nu, t)$ and 
$\tilde{b}(y,t)= b(y \, t^\nu, t)$.

First we assume that $\lambda \eta/(1 +\eta)>1/2$ which implies 
$\nu=1/2$ and consider the above equation
with $a$ and $b$ given by (\ref{defa}) and (\ref{defb}) with the
choice $\nu=1/2$. Neglecting all terms in Eq.~(\ref{ap1eqd}) 
which are of higher order than $1/t$ one obtains
\begin{equation}
\frac{\partial \tilde{P}}{\partial t}=
\frac{1}{2t} \left( \frac{\partial (y  \tilde{ P})}{\partial y}
+\frac{\partial^2 ( \tilde{P})}{\partial y^2} \right) \,.
\label{ap2eqd}
\end{equation}
This equation has a stationary solution 
$\tilde{P}= \frac{1}{\sqrt{2 \pi}} e^{-\frac{y^2}{2}}$ 
which implies that the core of the distribution $P(x,t)$ is 
Gaussian, $P(x,t)= \frac{1}{\sqrt{2 \pi \, t} } \, e^{-\frac{x^2}{2t}}$.

Second we assume that $\frac{\lambda \eta}{1 + \eta} < 1/2$ 
which implies $\nu=\frac{\lambda \eta}{1 + \eta}$.  
Neglecting all terms in Eq.~(\ref{ap1eqd}) 
which are of higher order than $1/t^{2\nu}$ 
one obtains
\begin{equation}
\frac{\partial \tilde{P}}{\partial t}=
\frac{1}{2t^{2\nu }} \left( 
{\rm sign}(y)\frac{\partial ( |y|^\eta \tilde{P})}{\partial y}
+\frac{\partial^2 ( \tilde{P})}{\partial y^2} \right) 
\end{equation}
Also this equation has a stationary solution
$\tilde{P}= c e^{-\frac{|y|^{\eta+1}}{\eta+1}}$ 
where $c= (\eta+1)^{\eta/(\eta+1)}/ 2 \Gamma(1/(\eta+1))$,
being $\Gamma(\cdot)$ the gamma function.
This implies that the core of the distribution $P(x,t)$ is 
\begin{equation}
P(x,t)= \frac{c}{t^{\nu}}  \, e^{-\frac{1}{\eta+1}
\left|\frac{x}{t^\nu} \right|^{\eta+1}}\,.
\label{predP}
\end{equation}
In conclusion for large times, the distribution approaches a Gaussian
when $\lambda \eta/(1 + \eta)>1/2$ and $\nu=1/2$, and approaches
the above anomalous density when $\nu=\lambda \eta/(1 + \eta)<1/2$.
Let us remark that the transition between these two densities
at $\lambda \eta/(1 + \eta) =1/2 $ is discontinuous, unless
$\eta=1$.

In Fig.~\ref{Fig_5} we show the rescaled probability distribution
associate to the process (\ref{xxx}) for $\lambda=0.55$ $\eta=5.0$
and different times together with the prediction of Eq.~(\ref{predP}).
The agreement is very good.

Let us remark that in the $\eta \to \infty$ limit we obtain
the same results already discussed in section~\ref{sec:3}
for the model with truncated tails. In particular for $\lambda < 1/2$
one has that the anomalous distribution (\ref{predP}) is constant 
in the interval $-t^\nu \le x \le t^\nu$ and vanishes elsewhere,
while for $\lambda > 1/2$ the core of the distribution is Gaussian.
The effect on tails when $\eta=\infty$ is simply a sharp cut 
at $|x|= t^\lambda$ which obviously disappears when time diverges.

Finally, in the limit $\eta \to 0$, the density (\ref{predP}) is
independent from time with value $P(x,t)\sim e^{-|x|}$, confirming
what obtained directly choosing $\eta=0$. Nevertheless, the
exponential decay is different, considering that for $\eta=0$ we found 
$P(x,t)\sim 2^{-|x|}$.  This difference is a consequence of the fact that the
large time limit and the vanishing $\eta$ limit cannot be interchanged
as it can be easily checked.

Now we are ready to discuss the front propagation behavior when a
reaction term is involved. The argument in Eq.s~(\ref{approx}), 
(\ref{thetaP2}) and (\ref{xf2}) would give $x_F \simeq \sqrt{2}\, t$ 
in the Gaussian case (i.e., when $\lambda \eta/(1 + \eta)>1/2$) and 
$x_F \simeq (1+\eta)^{1/(1+\eta)}\, t^\delta$ with
\begin{equation}
\delta = \frac{1 + \lambda \eta}{1 + \eta}\,.
\label{scalingdelta}
\end{equation}
for distribution (\ref{predP}) (i.e., when $\lambda \eta / (1 +
\eta)<1/2$). These two scalings can be observed only for a transient
time since, as we will discuss below, the correct asymptotic scaling
is $x_F (t) \sim t^\lambda$, as shown in Fig.~\ref{Fig_6} where the
front position is shown for $\lambda=0.55$, with reaction parameter
$\tau=1$ and different values of $\eta$.  In all cases the asymptotic
scaling is $x_F(t) \sim t^\delta$ with $\delta=\lambda=0.55$, but
there is a transient in which the exponent is given by
Eq.~(\ref{scalingdelta}). In the inset it is clearly shown for
$\eta=1$ the preasymptotic and asymptotic behavior of the front
position.

A simple theoretical explanation to justify the behavior of the
asymptotic front properties goes as follows. Let us
first remark that the forward jump probability (\ref{xpx}) when $|x|$
is of the order of $t^{\lambda + \epsilon}$ vanishes as $t^{-\epsilon
\eta}$, therefore, for large times the support of the distribution
$P(x,t)$ vanishes when $|x| \sim t^{\lambda + \epsilon}$.  This
implies that the front $x_F (t)$ cannot move faster then $x_F (t) \sim
t^\lambda$ when $t$ is large. On the other hand, when 
$|x|\sim t^{\lambda - \epsilon}$, the forward jump probability (\ref{xpx})
tends to $1/2$ for large times for any positive $\epsilon$, this
implies that the front $x_F (t)$ moves at least as fast as $x_F (t)
\sim t^\lambda$ when $t$ is large. In conclusion, $x_F (t) \sim t^\lambda$. 

A rough determination of the crossover time, i.e. the time at which
the front propagation starts to reach its asymptotic behaviour, can be
given with a matching between the position of the front in the
preasymptotic regime, $x_F(t)\simeq (1+\eta)^{1/(1+\eta)} t^\delta$, and
the value $\tilde x(t)$ at which the jump probability~(\ref{xpx}) is
small, $p(\tilde x,t)=1/(2+(|\tilde x|/t^\lambda)^\eta)=
1/(2+A)\ll 1$, i.e. when $\tilde x(t)\simeq A^{1/\eta}t^\lambda$, 
where $A$ is an appropriately large constant. 
The matching between those behaviours gives
\begin{equation}
t^*\simeq \left(\frac{A^{1/\eta}}{(\eta + 1)^{(\eta + 1)}}
\right)^\frac{(1+\eta)}{(1-\lambda)}.
\label{eq:match}
\end{equation}
With $A=8$, corresponding to a threshold $1/10$ for the jump
probability~(\ref{xpx}), one has a good agreement with the actual
results, as shown in the insets of figure~\ref{Fig_6}.

\section{Conclusions\label{sec:5}}
We study a class of modified random walk processes, whose probability
distributions can be analytically determined, which can have, at
varying a control parameter, standard or sub-diffusive behavior.  For
reaction-diffusion systems, with a pulled reaction function, the
scaling exponent of the diffusive problem is not relevant for the
asymptotic behavior of the front: the basic ingredient to determine
the front propagation behavior is the shape of tails of the
probability distribution.  This holds both for standard and
sub-diffusion.

Our results show, as already discussed
in~\cite{Sokolov2006,Mendez2009,Nepomnyashchy2016}, that the details
of the diffusion and reaction dynamics play a fundamental role 
in determining the front propagation features.

We conclude mentioning some topics which would be interesting to
investigate.  In our study we use a macroscopic description in terms of
concentration, in addition we considered a pulled reaction function
which acts even for very small concentration.  The macroscopic approach
is not appropriate if the density of transported ``particles'' is not
very large and therefore the discrete nature of the population cannot
be neglected.  We expect that the discreteness of the population will
impact in nontrivial ways on the spatial propagation properties of the
population.  In addition, a topic to investigate, even in the
macroscopic approach, is the role of shape of $f(\theta)$. For instance
we can probe how the use of reaction term in
the class of Allee non linearity (see for example~\cite{neufeld2010})
impact on the front propagation behavior.

\section*{Acknowledgments}
We thank the GSSI (L'Aquila) for organizing the workshop 
``Non standard transport NsT@GSSI'' (July 2015) in which this work began.
\bibliographystyle{model1-num-names}

\end{document}